\documentclass[preprint,aps,notitlepage, nofootinbib,superscriptaddress]{revtex4-1}
\RequirePackage{xspace} 
\usepackage{amsmath,amssymb}
\usepackage{mathrsfs}
\usepackage{bm}
\usepackage{algpseudocode,algorithm}

\usepackage{graphicx} %
\usepackage{color} %
\usepackage{cases}
\usepackage[colorlinks=true, pdfstartview=FitV, linkcolor=red, citecolor=blue, urlcolor=blue]{hyperref}
\usepackage[utf8]{inputenc}
\usepackage[T1]{fontenc}
\usepackage[english]{babel}
\usepackage[caption=false]{subfig}

\graphicspath{{./figures/}}

\newcommand{\diff}{\mathrm{d}}

\newcommand{\p}{\partial}

\newcommand{\e}{\mathrm{e}}

\newcommand{\calZ}{\mathcal{Z}}

\newcommand{\mc}{\mathrm{c}}
\newcommand{\mh}{\mathrm{h}}
\newcommand{\mo}{\mathrm{o}}

\newcommand{\mt}{\mathrm{t}}

\newcommand{\blue}[1]{{\color{blue}{#1}}}

\DeclareMathOperator*{\argmax}{arg\,max}

\allowdisplaybreaks[1]
\usepackage[normalem]{ulem}  

\newcommand{\bs}{\boldsymbol}
\begin{document}
%
%
\title{Phase transition encoded in neural network}
\preprint{RBRC-1301}
\author{Kouji Kashiwa}
\email[]{kashiwa@fit.ac.jp}
\affiliation{Fukuoka Institute of Technology, Wajiro, Fukuoka 811-0295, Japan}

\author{Yuta Kikuchi}
\email[]{yuta.kikuchi@riken.jp}
\affiliation{RIKEN BNL Research center, Brookhaven National Laboratory, 
Upton, NY, 11973, USA}

\author{Akio Tomiya}
\email[]{akio.tomiya@riken.jp}
\affiliation{RIKEN BNL Research center, Brookhaven National Laboratory, 
Upton, NY, 11973, USA}
\begin{abstract}
We discuss an aspect of neural networks for the purpose of phase
 transition detection.
To this end, we first train the neural network by feeding Ising/Potts
 configurations with labels of temperature so that it can
 predict the temperature of input. We do not explicitly supervise
 whether the configurations are in ordered/disordered phase. Nevertheless, we
 can identify the critical temperature from the parameters (weights
 and biases) of trained neural network.
We attempt to understand how temperature-supervised neural networks 
 capture the information of phase transition by paying attention to what
 quantities they learn.
Our detailed analyses reveal that they learn different physical
 quantities depending on how well they are trained.
Main observation in this study is how the weights in the
 trained neural-network can have information of the phase transition in
 addition to temperature.
\end{abstract}

\maketitle

\newpage
\section{Introduction}\label{sec:introduction}

Exploring phases of matters is one of the most important tasks to reveal infrared structure of the underlying microscopic physical system. Their phases are classified based on the symmetries that the theory possesses~\cite{landau1937theory,ginzburg1950theory}.
In a theory with several distinct phases, phase transitions occur at their boundaries. Among them, the nature of second order phase transition is solely determined by number of dimensions, global symmetries of underlying theory independent of its microscopic details, i.e., classified by the universality class.
In reality, however, analytic determination of various phases or detection of phase transitions based on the data of microscopic theory is generally a hard problem because we mostly find it difficult to exactly solve them or identify the corresponding infrared theory. Therefore, tremendous amount of works have been devoted to unravel them with numerical approaches.
An obvious and major obstacle is that the larger the number of degrees of freedom grows the harder the numerical analyses become.

Machine learning has grown up rapidly in the field of computer science and made prominent successes in pattern recognition, image processing, etc.
Recently, we have witnessed that  the machine learning has also been applied in various branches of physics. 
Detection of phase transitions is one of the intriguing examples that machine learning may make new progresses, and several approaches have already been proposed and examined in simple models such as spin systems. In those works, the trainings are carried out with supervision~\cite{carrasquilla2017machine,Tanaka:2016rtu,ohtsuki2016deep,mano2017phase,schindler2017probing,broecker2017machine,ch2017machine,zhang2017quantum,ponte2017kernel,zhang2017machine,arai2018deep,beach2018machine,Wetzel:2017ooo,richter2018machine,Suchsland2018,Kim2018} or without one~\cite{wang2016discovering,morningstar2017deep,van2017learning,hu2017discovering,wetzel2017unsupervised,Iso:2018yqu}. Here, the input data is prepared independently of the neural network or its training process. For instance, the Monte Carlo simulation or experimental data based on the physical system of our interest provide the necessary input data.

One approach to detect a phase boundary of a physical system is a supervised binary classification, where a neural network is trained so that it can distinguish its ordered and disordered phases. Indeed, it reasonably detects the phase transition in several models from their raw data~\cite{carrasquilla2017machine}.
More recently, some implicit connections between latent variables and physical quantities have been extensively studied toward the ultimate goal of eliminating black-box nature of machine learning technique \cite{carrasquilla2017machine,Wetzel:2017ooo,zhang2017machine,Kim2018}.
It also succeeded in detecting non-standard phase transition such as topological phase transitions and BKT phase transition with the help of feature engineerings based on physical insights.~\cite{zhang2017quantum,zhang2017machine,beach2018machine,richter2018machine,Zhang2018,Sun2018}.

Another intriguing approach was proposed in~\cite{Tanaka:2016rtu} to detect the phase transition by speculating that the information of order parameters are encoded in the weights of neural network as a consequence of training. 
They attempted to identify the critical temperature of the two dimensional Ising model based on supervised machine learning. A fully-connected as well as convolutional neural network are trained in such a way that it can correctly predict the target temperature of input spin configuration.
It is surprising that they succeeded in extracting the phase transition temperature from its weight because they did not feed any direct information about phase transition during the training.
It implies that the network \textit{spontaneously captures the phase transition} and encodes it in the machine parameters along the process of supervised learning of temperature although the underlying mechanism of the outcome was remained unresolved.

The purpose of this article is to understand the mechanism of this phase transition detection.
Understanding what this approach captures will be useful when we try to apply it to other unknown systems because it may not detect the phase transition correctly if it is triggered by different physical mechanisms such as quasi long-range order (BKT transition) or topology.
Indeed, it turns out that the method captures different physical quantities, i.e., the features of input configurations, depending on how we train the neural network. 
In order to illustrate the idea, we pay attention to simplified neural network architectures embodying the essence of temperature prediction.

\section{Critical temperature prediction}\label{sec:method_results}

We consider two dimensional Ising model, described by the Hamiltonian
$H(\{\sigma_i\}) = -J\sum_{\langle i,j\rangle}\sigma_i\sigma_j$,
where the coupling constant $J$ is taken to be positive. $\sigma_i\in\{-1,1\}$ represents a spin degree of freedom living on a site of a square lattice of size $L\times L$. We impose periodic boundary condition on the spin variables.
The sum is taken over nearest neighboring sites.
We redefine the Hamiltonian as
\begin{align}
 H(\{\sigma_i\}) = -\sum_{\langle i,j\rangle}\sigma_i\sigma_j,
\end{align}
by absorbing the coupling constant into the inverse temperature $K:=J/k_\text{B}T$ on the Boltzmann weight $e^{-KH}/{\cal Z}$, where $\calZ = \sum_{\{\sigma_i\}}\e^{-K H(\{\sigma_i\})}$ is the partition function.

We attempt to detect its critical temperature associated with the second-order phase transition.
Nevertheless, our first step is to construct a {\it thermometer} by employing a supervised feed-forward neural network. 
Then, we will examine the weights and biases of the trained neural networks, and attempt to identify the critical temperatures of two dimensional Ising model and 3-state Potts model. 

We generate 2000 Ising spin configurations at each temperature by employing the Metropolis-Hasting algorithm. 
They are fed to a neural network along with the 20 target temperatures.
We implement two types of neural network architectures by using KERAS package with TENSORFLOW as the backend: fully-connected and convolutional neural networks.

The former consists of fully-connected hidden and output layers as follows:
\begin{align}
\label{eq:FCNN}
\hspace{-.5cm}
\left[ \begin{array}{l}
\mathcal{I}=
\Big\{ \{ \sigma_{i} \} \Big| \text{ Ising configs on }L\times L \text{ lattice.} \Big\}
\\
 \downarrow
\
\left\{ \begin{array}{ll}
\text{Fully-connected (Dense) layer}  & \\
\text{Softmax activation} & \\
\end{array} \right.
\\
x_a\in\text{[0,1]}^{N_\mh} \text{ : hidden units}
\\
\downarrow
\ 
\left\{ \begin{array}{ll}
\text{Fully-connected (Dense) layer} & \\
\text{Softmax activation} & \\
\end{array} \right.
\\
y_K\in\text{[0,1]}^{N_\mo} \text{ : output}
\end{array} \right]
\end{align}
Let us describe it in detail here.
We denote input degrees of freedom by $\{\sigma_i\}\, (i=1,\dots,L\times L)$, which would be spins in case of Ising model.
A hidden unit $x_a\ (a=1,\dots,N_\mh)$ is given by,
\begin{align}
\label{eq:first_layer}
 x_a = \text{softmax}(w^{(1)}_{a i}\sigma_i+b^{(1)}_a)
 := \frac{\e^{w^{(1)}_{a i}\sigma_i+b^{(1)}_a}}{\sum_{a}\e^{w^{(1)}_{a i}\sigma_i+b^{(1)}_a}},
\end{align}
where repeated indices are summed. $w^{(1)}_{a i}$ and $b^{(1)}_a$ are weights and biases of the first layer, respectively.
In terms of weights $w^{(2)}_{\alpha a}$ and biases $b^{(2)}_\alpha$ of the second layer, variables $y_K\ (K=1,\dots,N_\mo)$ of output layer takes the same form as the hidden variables,
\begin{align}
 \label{eq:y_K}
  y_K = \text{softmax}(w^{(2)}_{K a} x_a+b^{(2)}_K).
\end{align}
Based on the output $\{y_K\}$, the temperature of input configuration is determined via
\begin{align}
\label{eq:K_predict}
 K^\mathrm{output} \equiv \argmax_{K}(y_K),
\end{align}
namely, the temperature $\alpha$ with the highest ``probability'' $y_K$ is picked as the output temperature.
The training is implemented by tuning the weights and biases with the
Adam optimizer~\cite{kingma2014adam} and the error function,
\begin{align}
 E(y_K,\bs{1}_{K= K^\text{target}_i})=-\frac{1}{N_\mo}\sum_i \bs{1}_{K=K^\text{target}_i}\ln y_K
\end{align}
which is the cross entropy between the target indicator function and output distribution function, where $i$ denotes the label of input configurations. The indicator function is defined by
\begin{align}
 \bs{1}_{K=K^\text{target}_i} = \left\{
 \begin{array}{ll}
 1 & (K = K_i^\text{target}) \\
 0 & (\text{otherwise})
 \end{array}
 \right.
\end{align}

The convolutional neural network consists of three convolutional layers followed by the final fully-connected layer~\eqref{eq:CNN}. We shall discuss it in more detail in Sec.~\ref{sec:energy_pred}.

Summary of our procedure is as follows:
\begin{description}
 \item[\bf Step 1] ~~Gather configurations via the standard Markov-chain Monte-Carlo
       method:
       We do not need the machine learning in this step.
       Data set $\{\sigma_i\}$ for each spin configuration at fixed
       $K$ is stored.
 \item[\bf Step 2]~~Train the neural-network as a thermometer: The input is the
       spin configuration and the output is the predicted temperature.\footnote{
       Because of the overlap of the energy probability
       distributions between each temperature, there is the upper-bound
       of accuracy of the prediction even if we ideally train the
       neural-network. See Appendix~\ref{sec:temperature_prediction} for detail.
       }
 \item[\bf Step 3]~~Analyze trained weights and biases appearing in the neural-network:
       We discuss how the machine parameters contain information of the phase transition.
\end{description}

\subsection{2D Ising model}

\begin{figure}[t]
\centering
\begin{minipage}{.49\textwidth}
\includegraphics[scale=0.5]{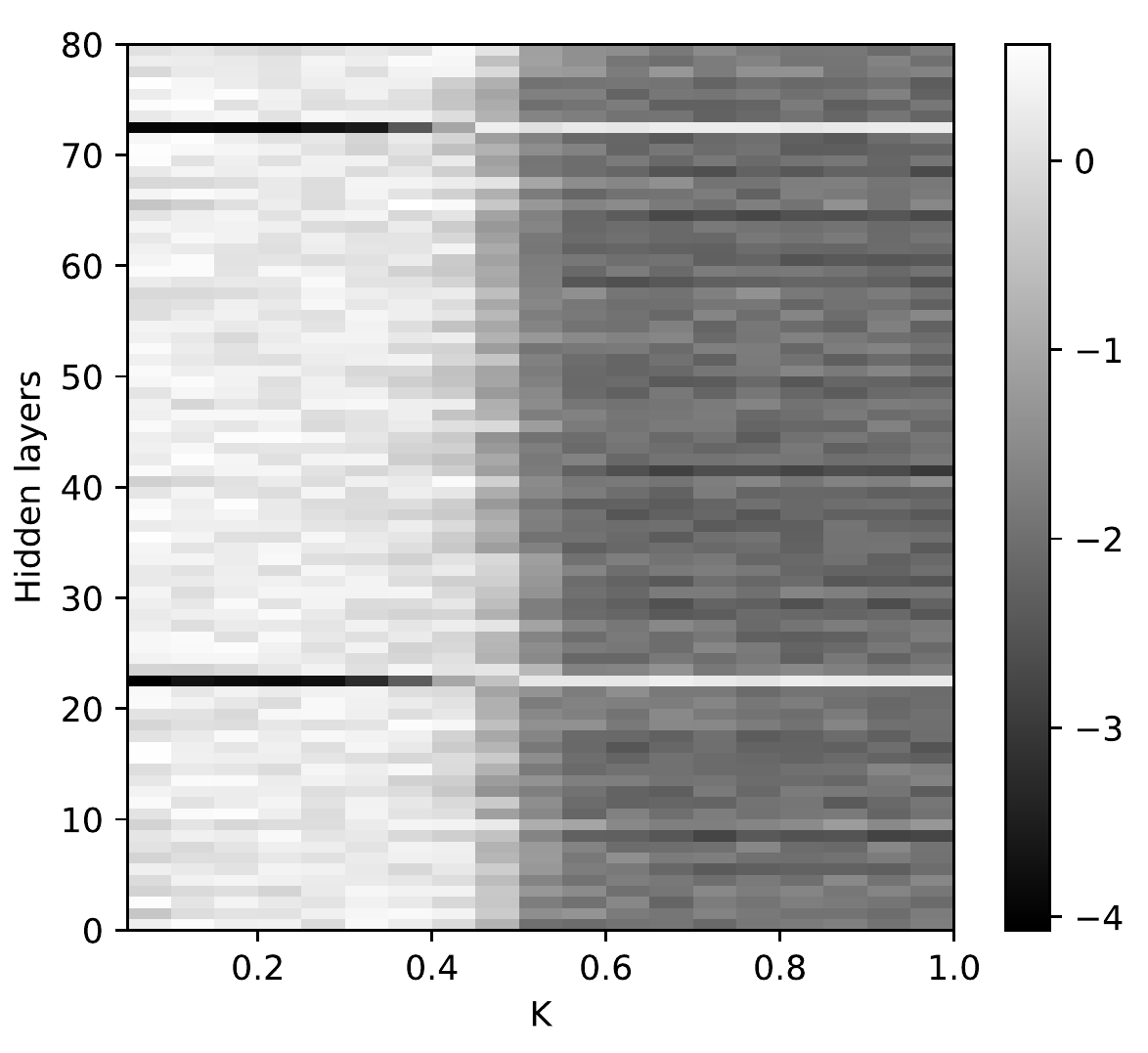}
\end{minipage}\
\begin{minipage}{.49\textwidth}
\includegraphics[scale=0.5]{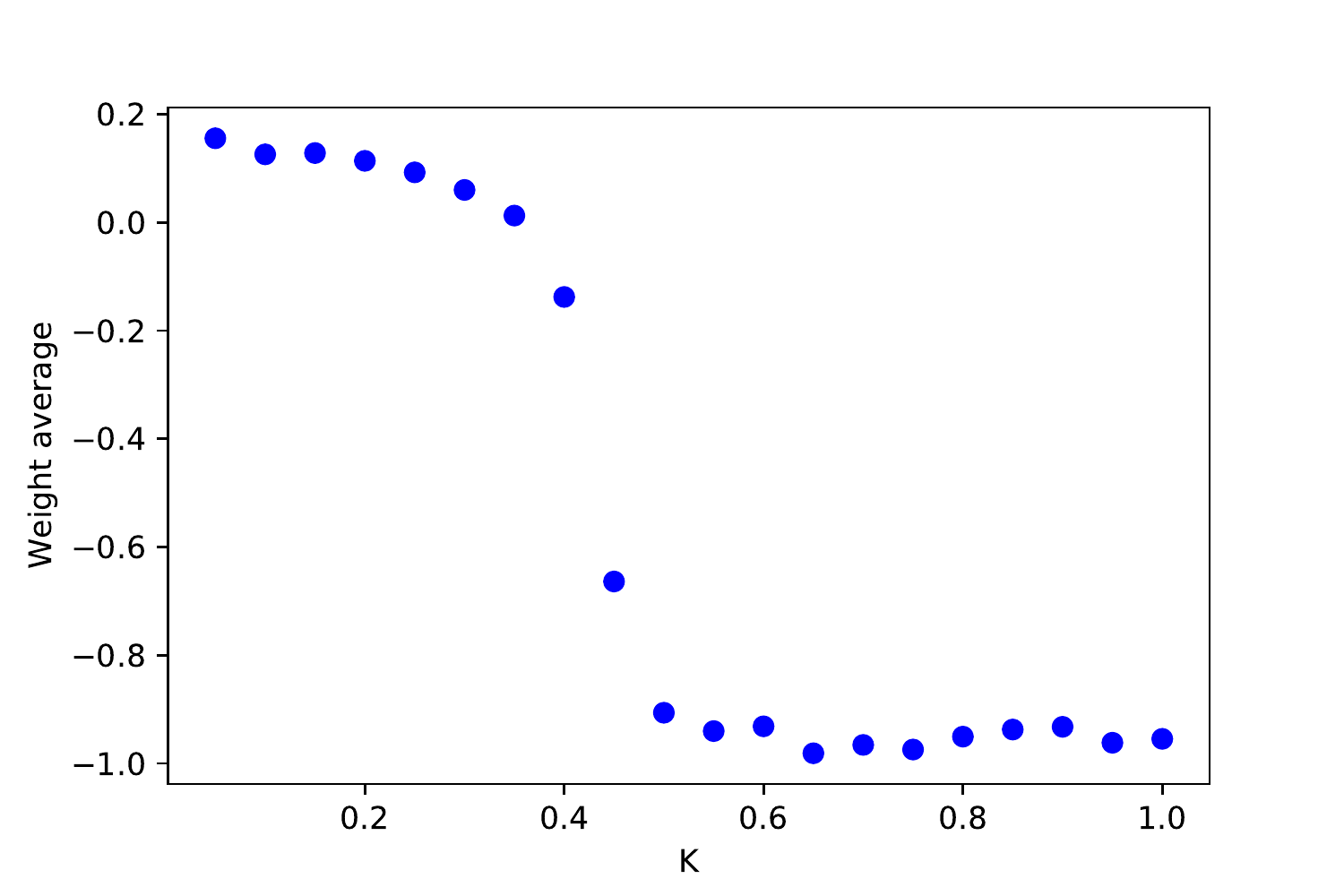}
\end{minipage}
\caption{The weights in the fully-connected layer and their average after the training in case of 2D Ising model. The horizontal axes represent the temperatures $K$ of input configurations. The vertical axis in the left panel corresponds to components connected to hidden units in the neural network. The vertical axis in the right panel shows average of weights for each $K$. The average value of weights significantly changes around the exact critical temperature $K^\text{exact}_\text{c}\simeq 0.4407$.}
\label{fig:weight_Ising}
\end{figure}

We briefly discuss how the phase transition detection works. We first
take a look at 2D Ising model, which was already studied in Ref.~\cite{Tanaka:2016rtu} with 100 target temperatures and its weights behave like an order parameter, i.e. spontaneous magnetization. Since our primary purpose is to understand its mechanism rather than to quantitatively estimate the critical temperature, we reduce the number of target temperatures to 20: $K=0.05,0.1,0.15,\dots,1.0$. 
We perform the supervised training using neural network \eqref{eq:FCNN}, with lattice size $L=16$, the number of hidden units $N_\mh=80$, and $N_\mo=20$ corresponding to the 20 target temperatures.
The critical temperature is exactly known to be $K^\text{exact}_\text{c}=\frac{1}{2}\ln(\sqrt{2}+1)\simeq 0.4407$~\cite{onsager1944crystal}.
The weights of second layer after the training is shown in Fig.~\ref{fig:weight_Ising}.
In Ref.~\cite{Tanaka:2016rtu}, the critical temperature was predicted by fitting the sum of the weights by $c_1\tanh[c_2(K-K_\text{c})]-c_3$ with free parameters $c_1,c_2,c_3$ for lattice sizes $L=8, 16, 32$.
Indeed, the average of the final weights appears to behave like an order parameter (right panel of Fig~\ref{fig:weight_Ising}).
We will discuss the detailed structure of the weights in the next section.



\subsection{2D 3-state Potts model}

Before getting into the detail of learning mechanism of critical
temperature of 2D Ising model, we take a look at another example, 2D
3-state Potts model.
The Hamiltonian is given by
\begin{align}
 H(\{\Phi_i\}) = -\sum_{\langle i,j\rangle}\delta(\Phi_i,\Phi_j),
\end{align}
where $\Phi_i$ takes three values, a generalization of Ising spin $\sigma_i$.
Hence, configurations $\{\Phi_i\}$ labeled by temperatures $K$ are the inputs of neural network.
The 2D 3-state Potts model is known to exhibit the second order phase transition at
$K_\text{c}\simeq 1.0050$ because of the
fluctuation unlike the simple Landau theory~\cite{Wu1982,baxter1973potts,alexander1975s}.

\begin{figure}[t]
\centering
\begin{minipage}{.49\textwidth}
\includegraphics[scale=0.5]{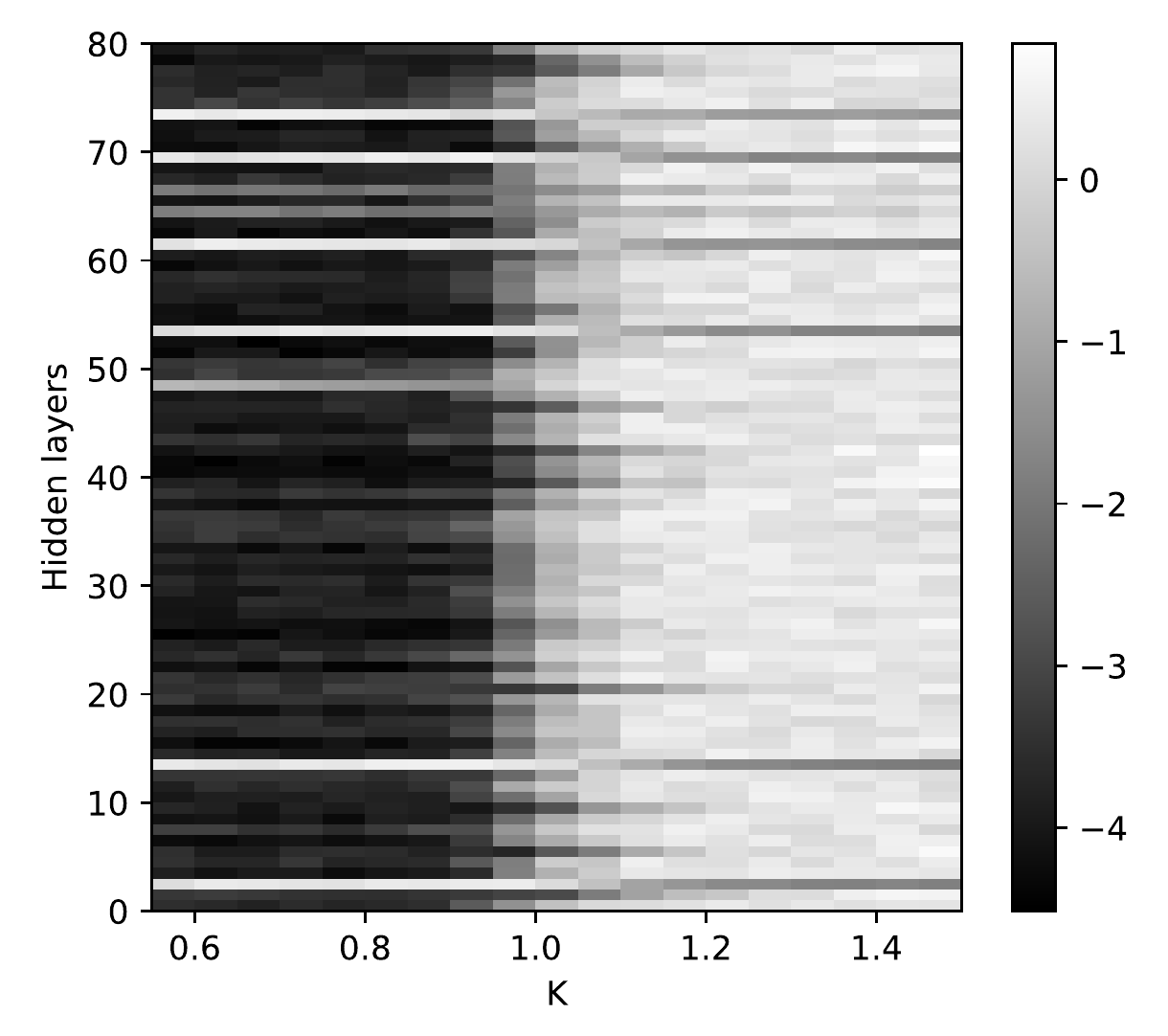}
\end{minipage}\
\begin{minipage}{.49\textwidth}
\includegraphics[scale=0.5]{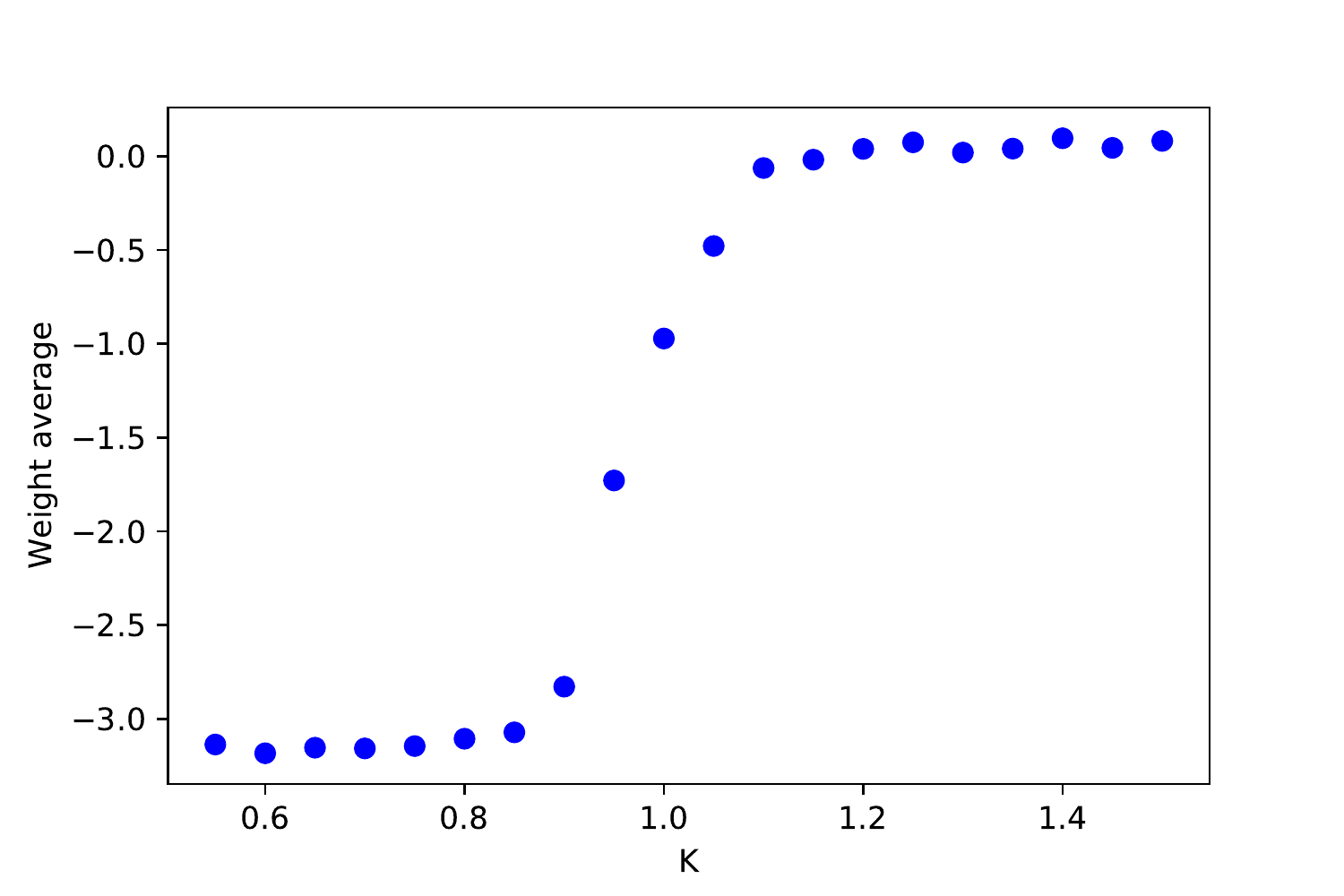}
\end{minipage}
\caption{The weights in the fully-connected layer and their average after the training in case of 2D 3-state Potts model (left panel). The average value of weights significantly changes around the critical temperature $K_\text{c}\simeq 1.0050$ (right panel).}
\label{fig:weight_Potts}
\end{figure}

After a training with the same neural network architecture as that for 2D Ising model, we obtain the weights and their average shown in Fig.~\ref{fig:weight_Potts}. We again find a drastic change in the weight structure around the critical temperature.

\section{Discussion}\label{sec:discussion}

So far, we have observed that the change in value of weights in the second layer signals the phase transition, which implies that the information of critical temperature is somehow encoded in the trained neural network. 
In what follows, we carefully examine the trained fully-connected/convolutional neural networks and attempt to understand what (physical) quantity they extract as a feature of input configurations and how it is related to temperature prediction as well as the critical temperature detection~\cite{Wetzel:2017ooo,Suchsland2018}.

\subsection{Magnetization encoded in neural network}

\begin{figure}[t]
\centering
\begin{minipage}{.49\textwidth}
\includegraphics[scale=0.5]{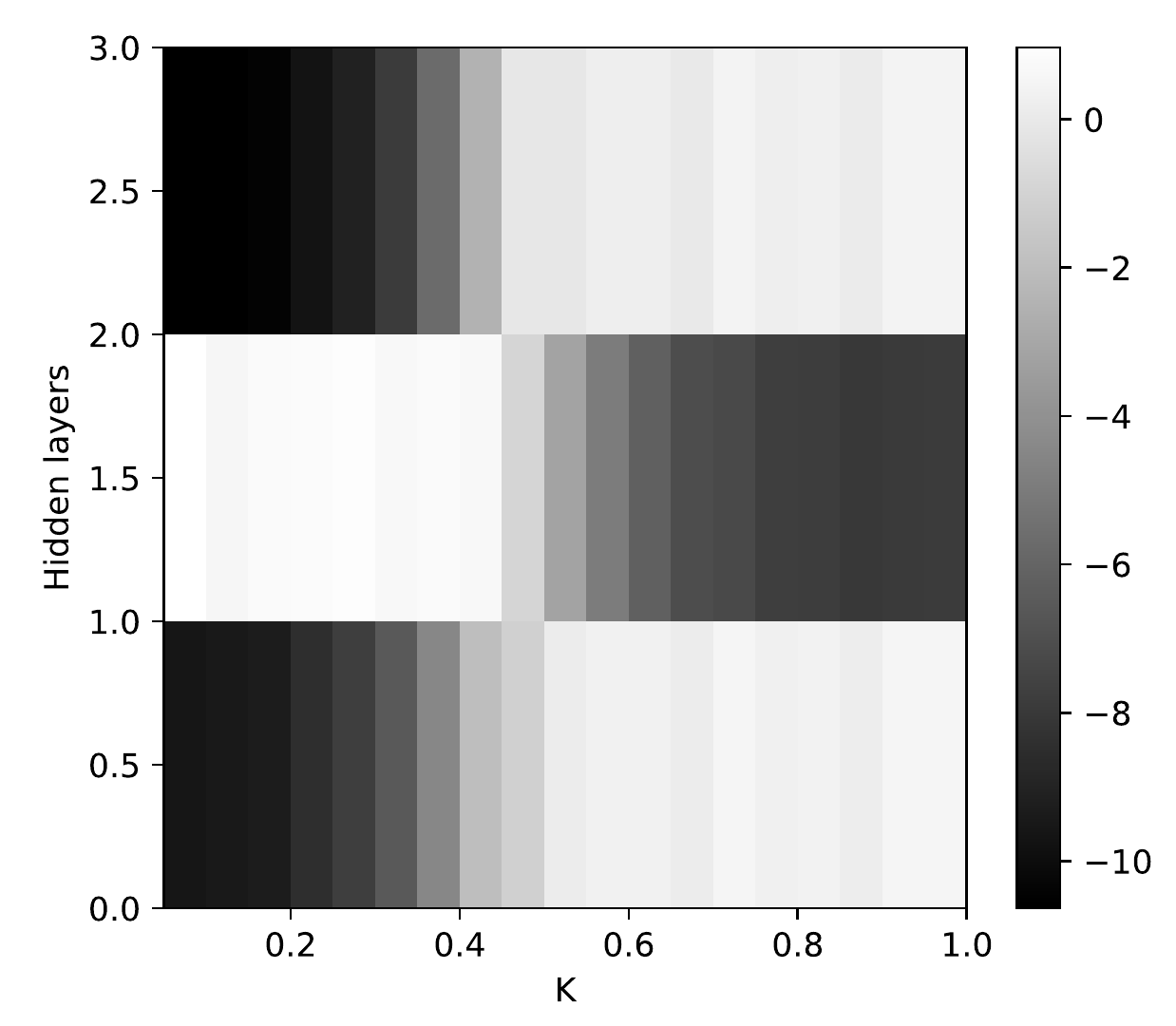}
\end{minipage}\
\begin{minipage}{.49\textwidth}
\includegraphics[scale=0.5]{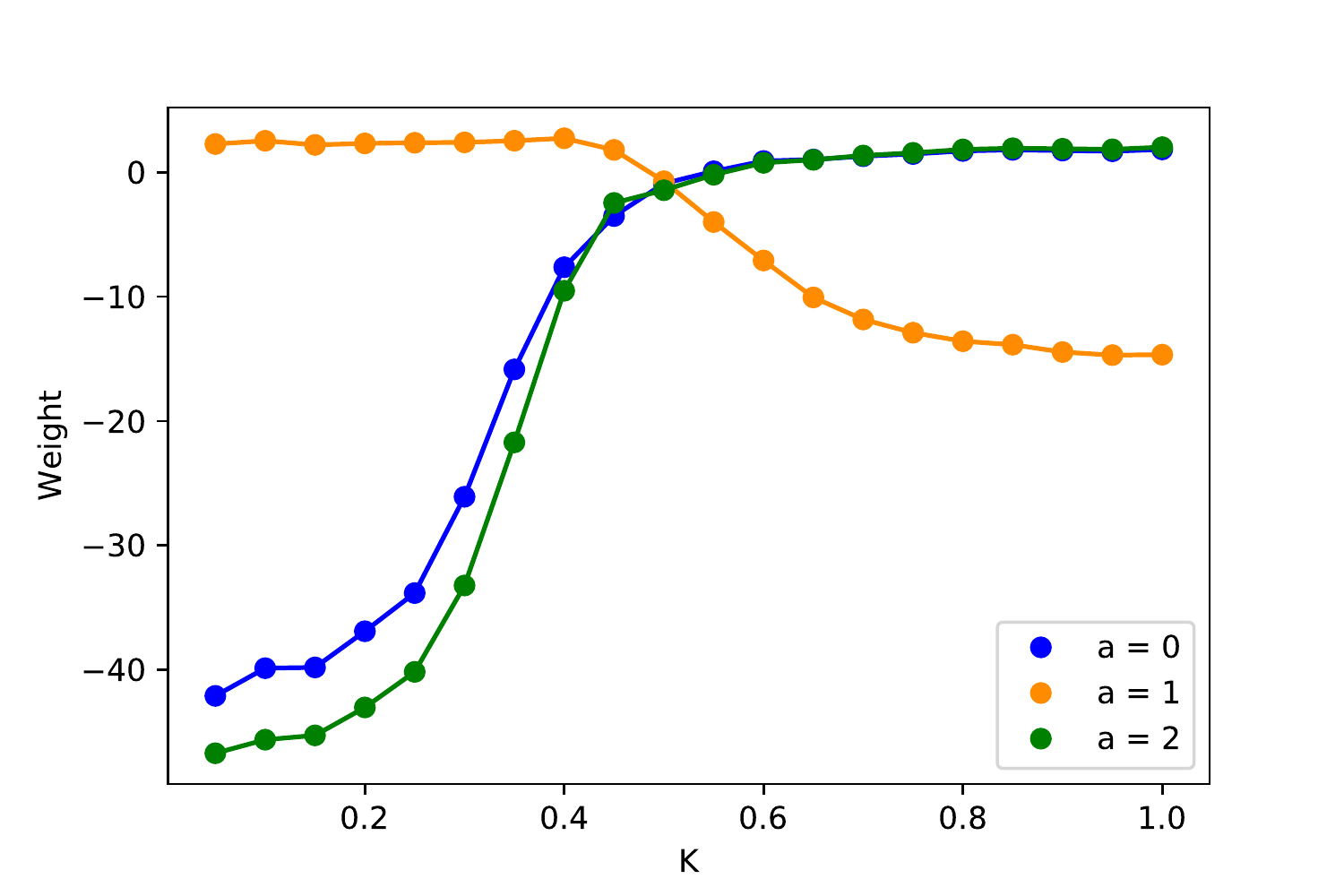}
\end{minipage}
\caption{The fully-connected weights $w^{(2)}_{Ka}$ of trained neural networks with three hidden units ($N_\mh=3$). The horizontal axes represent temperatures $K$ of input 2D Ising configurations. The structure change is still observed around critical temperature. One of the weights has an almost opposite temperature dependence, which also appeared in Fig.~\ref{fig:2DIsingDist}. See the main text for detailed discussion.}
\label{fig:layer_weight_hidden3}
\end{figure}

Since the order parameter of phase transition in the 2D Ising model is
the spontaneous magnetization, it sounds natural that it is encoded in
the neural network after the training.
To give a quantitative argument we construct a simplified model by
examining the weights and biases of training neural network in case of
2D Ising model.
First, we reduce the number of hidden unit in \eqref{eq:FCNN} from 80 to
3 for the sake of simplicity. We notice that it still captures
the critical temperature as shown in
Fig.~\ref{fig:layer_weight_hidden3}.

\begin{figure}[t]
\centering
\begin{minipage}{.49\textwidth}
\subfloat[Output of first layer]{
\includegraphics[scale=0.5]{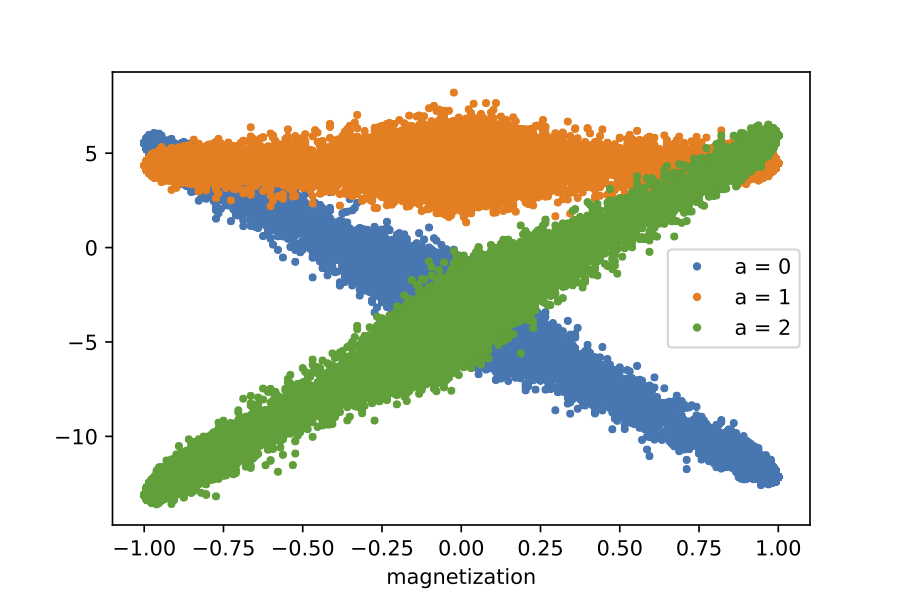}
\label{fig:mag_hidden3}
}\end{minipage}\
\begin{minipage}{.49\textwidth}
\subfloat[Parametrization in Eq.~\eqref{eq:cor_mag}]{
\includegraphics[scale=0.4]{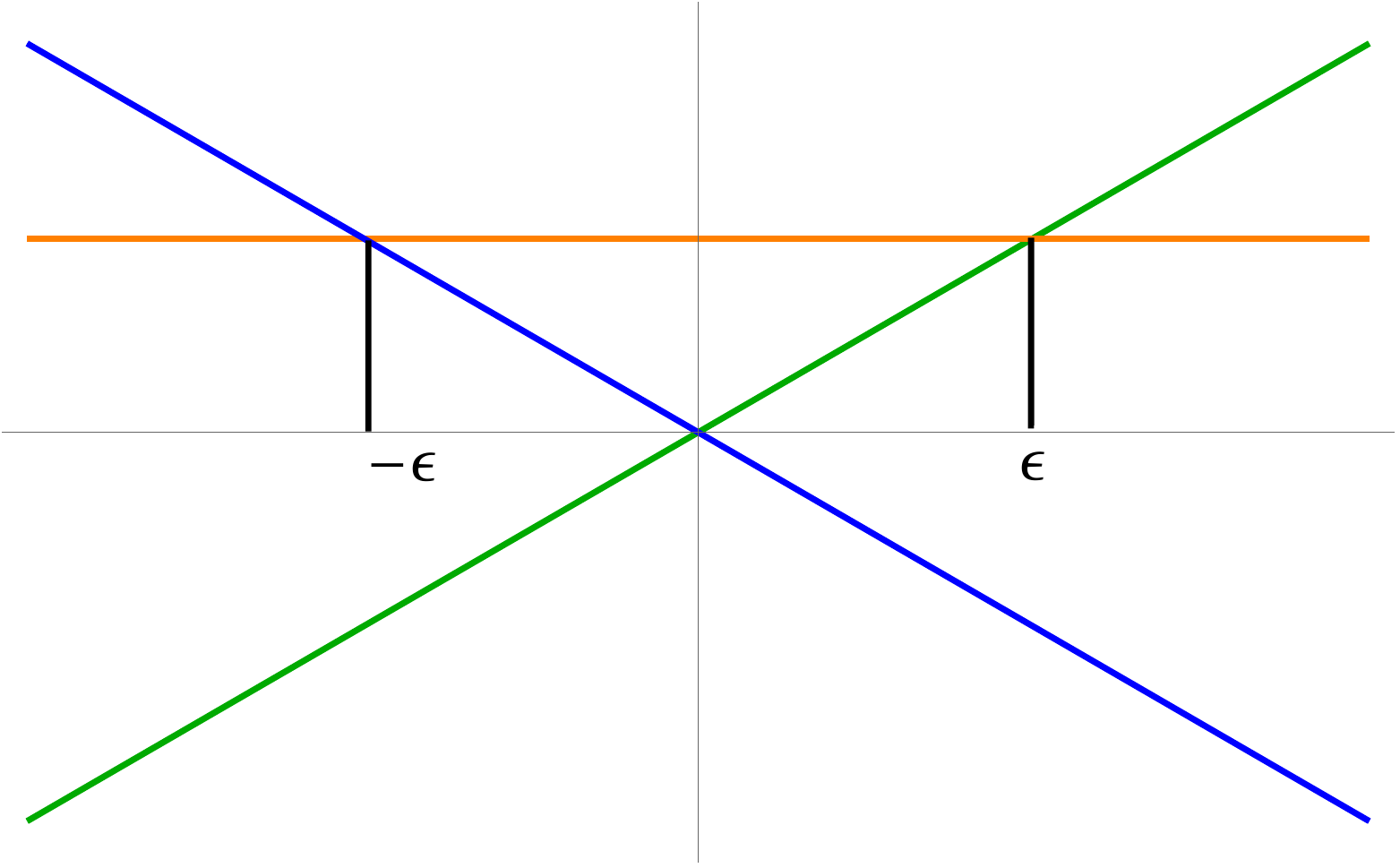}
\label{fig:simple_mag_hidden3}
}\end{minipage}
\caption{(a) Correlations between the output of the first layer and magnetization of input configuration.
The horizontal axis represents the magnetization per site of input Ising spin configuration and the vertical axis is $\tilde{x}_a=w^{(1)}_{a i}\sigma_i+b^{(1)}_a$.
(b) The model parametrization, where each line corresponds to each row of Eq.~\eqref{eq:cor_mag}.}
\label{fig:mag_correlation}
\end{figure}

Before modeling the second layer, we examine the characteristics of the
 first layer.
 Figure~\ref{fig:mag_correlation}
shows the correlation between the output of the first layer $\tilde{x}_a:=w^{(1)}_{a
 i}\sigma_i+b^{(1)}_a$ of Eq.~\eqref{eq:first_layer} and magnetization
 density of the input Ising spin configuration.
From this observation, we model these output by three lines linear in
 magnetization $m(\{\sigma\})$ as shown in Fig.~\ref{fig:simple_mag_hidden3}:
\begin{align}
\label{eq:cor_mag}
  \tilde{x} =
 \begin{pmatrix}
 \tilde{x}_0 \\ \tilde{x}_1 \\ \tilde{x}_2
 \end{pmatrix}
 =
 \begin{pmatrix}
 -m \\ \epsilon \\ m
 \end{pmatrix},
\end{align}
where $\epsilon>0$ is a constant.
Furthermore, as an activation function, we use the max function, that
assigns 1 to the maximum entry and 0 to the rest, instead of softmax for
our purpose; this replacement does not change our final result.
$x_a =\text{max}(\tilde{x}_a)$ yields the following vectors depending on magnetization of inputs $m$,
\begin{align}
 m<-\epsilon:\ x=
 \begin{pmatrix}
 1 \\ 0 \\ 0
 \end{pmatrix},
 \quad
 -\epsilon\le m<\epsilon:\ x=
 \begin{pmatrix}
 0 \\ 1 \\ 0
 \end{pmatrix},
 \quad
 \epsilon\le m:\ x=
 \begin{pmatrix}
 0 \\ 0 \\ 1
 \end{pmatrix}.
\end{align}
The parameter $\epsilon$ may be interpreted as a threshold magnetization separating the ferromagnetic and paramagnetic phases~\cite{carrasquilla2017machine}.

Having understood the magnetization dependence of the three hidden units, we next analyze the second layer.
The weights in the layer are given in Fig.~\ref{fig:layer_weight_hidden3}.
We divide the temperature into three pieces: low, critical, and high temperatures, respectively represented by the following vectors,
\begin{align}
\label{eq:3temperature}
\text{Low K : }
\begin{pmatrix}
 1 \\ 0 \\ 0
\end{pmatrix}, 
\quad
\text{Critical K : }
\begin{pmatrix}
 0 \\ 1 \\ 0
\end{pmatrix}, 
\quad
\text{High K : }
\begin{pmatrix}
 0 \\ 0 \\ 1
\end{pmatrix},
\end{align}
in $N_\mo$-dimensional output space.
This procedure effectively reduces the output dimension~$N_\mo$ to 3.
According to Fig.~\ref{fig:layer_weight_hidden3}, we parametrize the weights in the following way,
\begin{align}
w^{(2)}=
\begin{pmatrix}
 -\Delta & 0        & -\Delta 
 \\ 
 -\delta & -\delta & -\delta
 \\ 
 0         & -\Delta & 0
\end{pmatrix}, 
\end{align}
with $\Delta>\delta>0$. Elements of $(K\times a)$-matrix $w^{(2)}$ are the weights $w^{(2)}_{Ka}$. We neglect the biases as they are much smaller than the weights. Precise parametrization is not necessary for the following discussion.
Then, $y_K= \text{max}(\tilde{y}_K)= \text{max}(w^{(2)}_{Ka}x_a+b^{(2)}_K)$\footnote{Remember that the max function here is defined to be a function assignning 1 to the maximum entry and 0 to the rest.} yields the following output,
\begin{align}
 m<-\epsilon&:\
  y_K= \text{max}(w^{(2)}_{K0})
 = 
 \begin{pmatrix}
 0 \\ 0 \\ 1
 \end{pmatrix},
 \\
 -\epsilon\le m<\epsilon&:\ 
 y_K= \text{max}(w^{(2)}_{K1})
 =
 \begin{pmatrix}
 1 \\ 0 \\ 0
 \end{pmatrix},
 \\
 \epsilon\le m&:\ 
 y_K= \text{max}(w^{(2)}_{K2})
 =
 \begin{pmatrix}
 0 \\ 0 \\ 1
 \end{pmatrix},
\end{align}
which are predicted as low, high, and low temperature, respectively.
Since the configurations with $m<-\epsilon$ and $m>\epsilon$ are in the
ordered phase, they are correctly predicted as low temperature phase. 
The configurations with $-\epsilon\le m<\epsilon$ is also correctly
predicted as high temperature.
However, it cannot detect the intermediate temperature, i.e., the critical temperature in this case.
Furthermore, the different temperatures within the ordered/disordered
phase are not distinguished by the trained neural network even if we introduce more than three target temperatures in Eq.~\eqref{eq:3temperature} because upper branch of the weights (and biases)  are almost temperature independent above and below $K_\mc$, respectively, as seen in
Fig.~\ref{fig:layer_weight_hidden3}. 
Therefore, the trained network is capable of distinguishing only high or low temperature.

One might think that this is due to the fact that we have single threshold parameter $\epsilon$ corresponding to the three hidden unit. Actually, we can increase the number of threshold parameters by introducing more hidden units. But, it does not lead to higher resolution of temperatures. In fact, even if we increase the number of hidden units, the weights in the second layer show only two patterns of temperature dependence: most of them behave like blue and green curves and the rest behave like the orange curve in Fig.~\ref{fig:layer_weight_hidden3}. This is exactly what is observed in Fig.~\ref{fig:weight_Ising}.
Increasing the hidden units simply results in duplicating the second layer's weights that are already observed in  Fig.~\ref{fig:layer_weight_hidden3}, and consequently, the predicted temperature is either high or low no matter how many hidden units are introduced.

What we have learned from the above analyses is as follows:
The network obtained the information of magnetization which manifests itself in the output of the first layer. 
However, it is hard for this network to discriminate each temperature except for the difference between ordered and disordered phases based on the magnetization.
From the viewpoint of machine learning parameters, this is due to the fact that the weights of the second layer are temperature independent except around the critical temperature.

\subsection{Energy and temperature prediction}
\label{sec:energy_pred}

We have found in the last subsection that the magnetization of 2D Ising model was build in the temperature-supervised fully-connected neural network, that in turn allows us to read off the critical temperature from its weight structure.
While the critical temperature seems to be well detected, we have not discussed the temperature prediction itself.
Interestingly, the accuracy of temperature learning can be theoretically computed to be 40.1\% in case of 2D Ising model under the above setup, giving the upper bound for the accuracy of temperature prediction by machine learning~\cite{Aoki2018} (see Appendix~\ref{sec:temperature_prediction} for details). 
However, we note that the test accuracy of temperature prediction by fully-connected neural network is 16.8\%,
which is not even close to theoretically predicted accuracy 40.1\%.
We could train better from the viewpoint of temperature prediction although we did not need for the phase transition detection by extracting magnetization as we have already seen.
What happens if we design the neural network architecture in such a way that the temperature prediction accuracy improves?

To answer the question, we use a convolutional neural network shown below, enabling us to achieve higher accuracy in two dimensional image recognition.
\begin{align}
\label{eq:CNN}
\hspace{-.5cm}
\left[ \begin{array}{l}
\mathcal{I}=
\Big\{ \{ \sigma_{i} \} \Big| \text{ Ising configs on }L\times L \text{ lattice.} \Big\}
\\
 \downarrow
\
\left\{ \begin{array}{ll}
\text{Convolution}_{[(s_1,s_1) \text{-filter, } (s_1,s_1) \text{-stride, }C_1 \text{-channels}]} & \\
\text{ReLU activation} & \\
\text{Convolution}_{[(s_2,s_2) \text{-filter, } (s_2,s_2) \text{-stride, }C_2 \text{-channels}]} & \\
\text{ReLU activation} & \\
\text{Convolution}_{[(s_3,s_3) \text{-filter, } (s_3,s_3) \text{-stride, }C_3 \text{-channels}]} & \\
\text{ReLU activation} & \\
\text{Flatten} & \\
\end{array} \right.
\\
x_a\in\mathbb{R}^{N_\mh = \ L^2/(s_1s_2s_3)^2 \times C_3 } \text{ : hidden units}
\\
\downarrow
\ 
\left\{ \begin{array}{ll}
\text{Fully connected layer} & \\
\text{Softmax activation} & \\
\end{array} \right.
\\
y_\alpha\in\text{[0,1]}^{N_\mt}\text{ : output}
\end{array} \right]
\end{align}
The network has three convolutional layers (Conv2d layers from KERAS) with square filters with size $(s_i,s_i)$, strides $(s_i,s_i)$ and the number of channels $C_i$, each of which is followed by ReLU activation function. Then, the output is passed to fully-connected layer, whose input and output are of out interest and analyzed in detail.
The output of the second last layer, denoted by $x_a$, is given by
\begin{align}
 x_a = \text{ReLU}(\tilde{x}_a) = \text{Max}(\tilde{x}_a, 0),
\end{align}
takes a value in $[0,\infty)$.
$\tilde{x}_a$ is an output of the previous layer before passed to the ReLU activation.
Then, the output $x_a$ plays a role of input of the fully-connected layer to give a prediction of temperature via \eqref{eq:y_K} and \eqref{eq:K_predict}.

Following the last subsection, we put three hidden units playing a role of input of the fully-connected layer and attempt to understand what the neural network learns as a result of the training.
To this end, the parameters are set as follows:
\begin{align}
 L=16, \quad
 (s_1,s_2,s_3) = (2,2,4),\quad 
 (C_1,C_2,C_3) = (64,32,3),
\end{align}
leading to the number of hidden units $N_\mh = 3$.
With this setup we achieved 35.9\% of accuracy on a test data set. It is twice higher than fully-connected layers, although still not very close to the bound partially due to the lack of statistics and very small number of hidden units.

\begin{figure}[t]
\centering
\begin{minipage}{.49\textwidth}
\subfloat[Correlation with magnetization.]{
\includegraphics[scale=0.5]{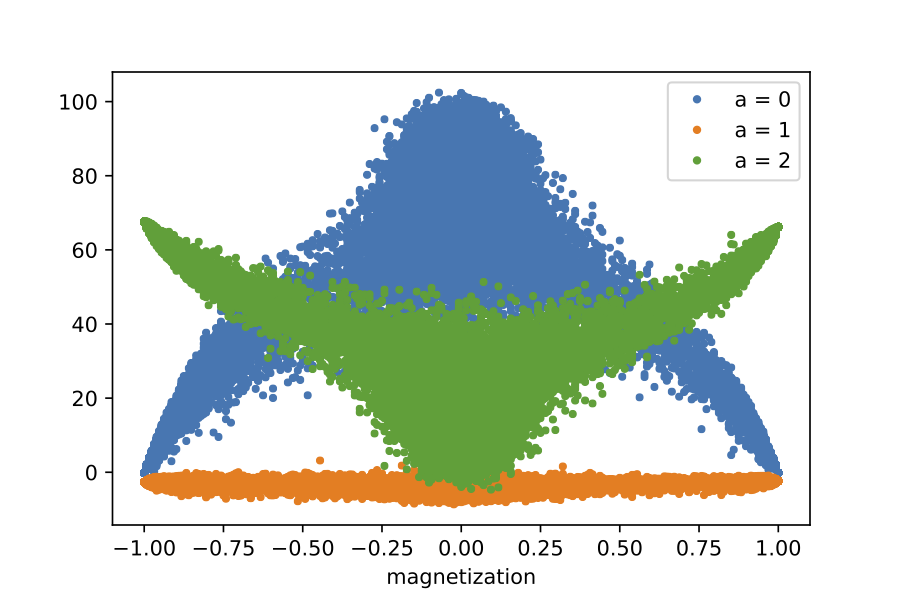}
\label{fig:hidden3_CNN_mag}
}\end{minipage}\
\begin{minipage}{.49\textwidth}
\subfloat[Correlation with internal energy.]{
\includegraphics[scale=0.5]{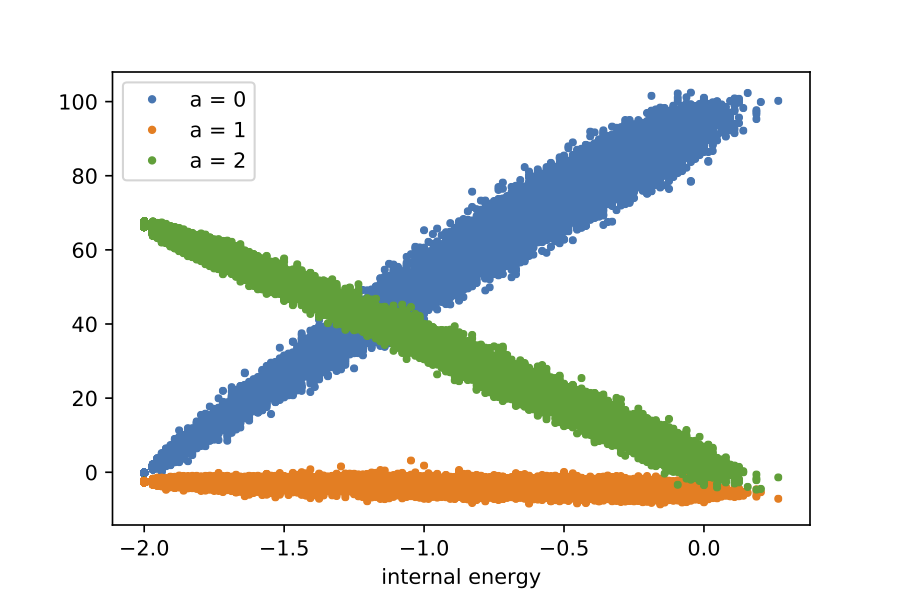}
\label{fig:hidden3_CNN_energy}
}\end{minipage}
\caption{Correlations between physical quantities and the weights of fully-connected layer in the trained neural network with five hidden units, each of which corresponds to vertical axis of the left and right panel. The horizontal axes in (a) and (b) show magnetization and internal energy of input 2D Ising configurations, respectively.}
\label{fig:hidden3_CNN}
\end{figure}

\begin{figure}[t]
\centering
\begin{minipage}{.49\textwidth}
\includegraphics[scale=0.5]{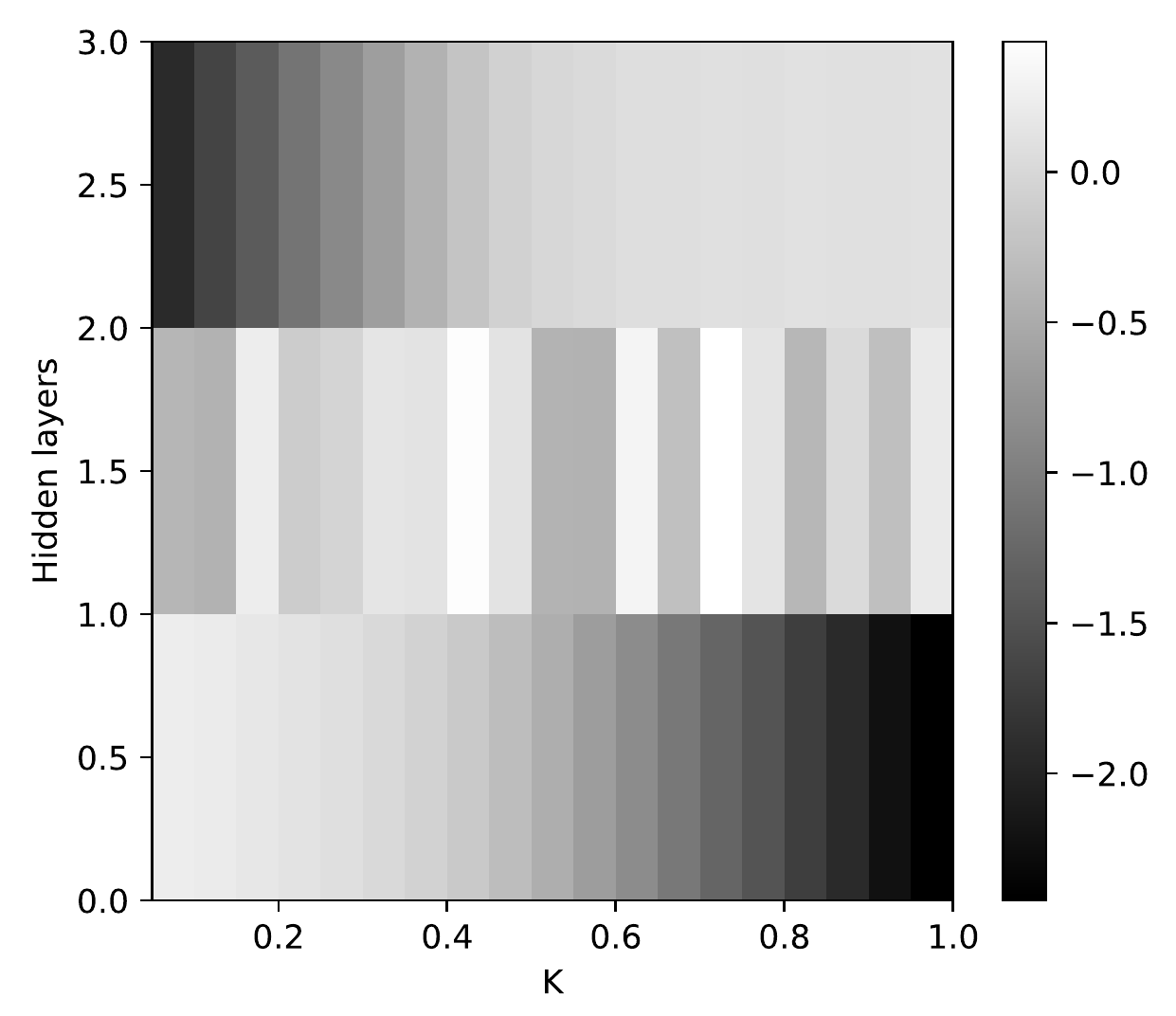}
\end{minipage}\
\begin{minipage}{.49\textwidth}
\includegraphics[scale=0.5]{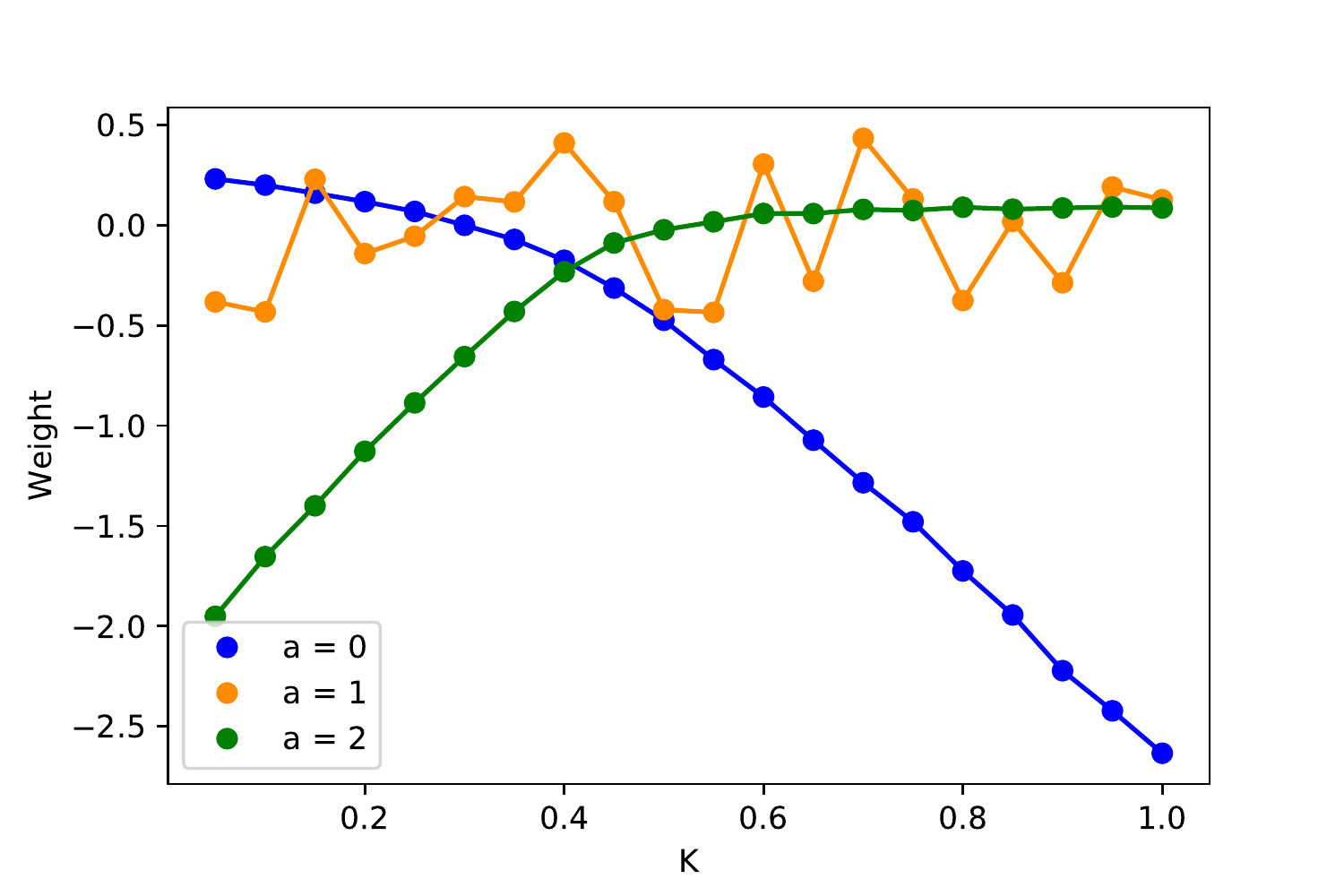}
\end{minipage}
\caption{The fully-connected weights of trained convolutional neural networks with three hidden units ($N_\mh=3$). The horizontal axes represent temperatures $K$ of the input 2D Ising configurations. The value of  the first and last weights gradually change as temperature increases in contrast to the previous case. See the main text for detailed discussion.}
\label{fig:weight_CNN}
\end{figure}

Figures~\ref{fig:hidden3_CNN} shows the correlation of the output of first layer with magnetization (Fig.~\ref{fig:hidden3_CNN_mag}) and internal energy (Fig.~\ref{fig:hidden3_CNN_energy}), respectively.
We clearly see the transition in what the neural network has learned. 
The first-layer outputs are now proportional to the internal energy rather than the magnetization of input configurations.
Also, we notice that the weights of the second layer obtain mild dependence on temperature, implying the information of the critical temperature is blurred (Fig.~\ref{fig:weight_CNN});
as we shall see momentarily, the oscillating orange line is irrelevant in the temperature prediction.

We now proceed to a simplified parametrization of the last two layers.
Here, $(w^{(1)}_{ai},b^{(1)}_{a})$ and $(w^{(2)}_{Ka},b^{(2)}_{K})$ stand for weights and biases in the second last and last layer, respectively.
As we have already mentioned and checked in
Fig.~\ref{fig:hidden3_CNN_energy}, the output of the first layer is
proportional to energy $E(\{\sigma_i\})$ of input configuration
$\{\sigma_i\}$. After passing to the ReLU activation,
\begin{align}
 x_a = \text{ReLU}(w^{(1)}_{ai}\sigma_i+b^{(1)}_a),~~~~
 x=
 \begin{pmatrix}
 x_0 \\ x_1 \\ x_2
 \end{pmatrix} =
 \begin{pmatrix}
 E+2\epsilon \\ 0 \\ -\phi E
 \end{pmatrix},
 \label{Sec:xa}
\end{align}
where $\epsilon$ and $\phi$ are positive constants. Domain of $E$ is restricted so that $x_a$ does not take a negative value. 

Having observed that the input of the fully-connected layer is proportional to $E$, let us consider what would be an optimal estimation of the temperature of an input configuration~$\{\sigma_i\}$~\cite{Aoki2018}.
The probability $P(\{ \sigma_i \};K)$ that the configuration $\{\sigma_i\}$ appears at temperature $K$ is given by
\begin{align}
 P(\{ \sigma_i \};K) = \frac{\e^{-KE(\{\sigma_i\})}}{\calZ(K)}.
\end{align}
Therefore, the likelihood that $\{\sigma_i\}$ is generated at temperature $K$ is represented by the following ``probability'',
\begin{align}
 \label{eq:theory_output}
 y_K^\text{theory}
 = \frac{P(\{ \sigma_i\} ;K)}{ \sum_{K'}P(\{ \sigma_i\} ;K')}
 =\text{softmax}(-KE+F),
\end{align}
where $F=-\ln\calZ(K)$ is the free energy and $K'$ is summed over the target temperatures.
Then, we obtain the estimated temperature,
\begin{align}
 K^\text{output} = \argmax_{K}\big[\text{softmax}(-KE+F))\big].
\end{align}
We remark here that the free energy is a function of temperature $K$ and holds the information of phase transition although it does not exhibit genuine singularity in finite systems.

Based on the above consideration, we guess fully-connected weights of our neural network so that the resultant output behaves like $\eqref{eq:theory_output}$. Then, we compare with the actual parameters of the trained network.
To this end, we parametrize the weights as follows:
\begin{align}
\label{eq:theory_weight}
w^{(2)}_{K0}=-\phi G(K) - pK + q, 
\quad
w^{(2)}_{K2}=-G(K) + rK + s,
\end{align} 
with constant parameters $p,q,r,s$ and a common nonlinear function
$G(K)$.
The orange curve in Fig.~\ref{fig:weight_CNN}, $w^{(2)}_{K1}$, is not relevant to our consideration because $x_1=\text{ReLU}(\tilde{x}_1)=0$.
We use an observation from the simulation to neglect the bias, where it is much smaller than the value of weights.

\begin{figure}[t]
\centering
\includegraphics[width=0.5\textwidth]{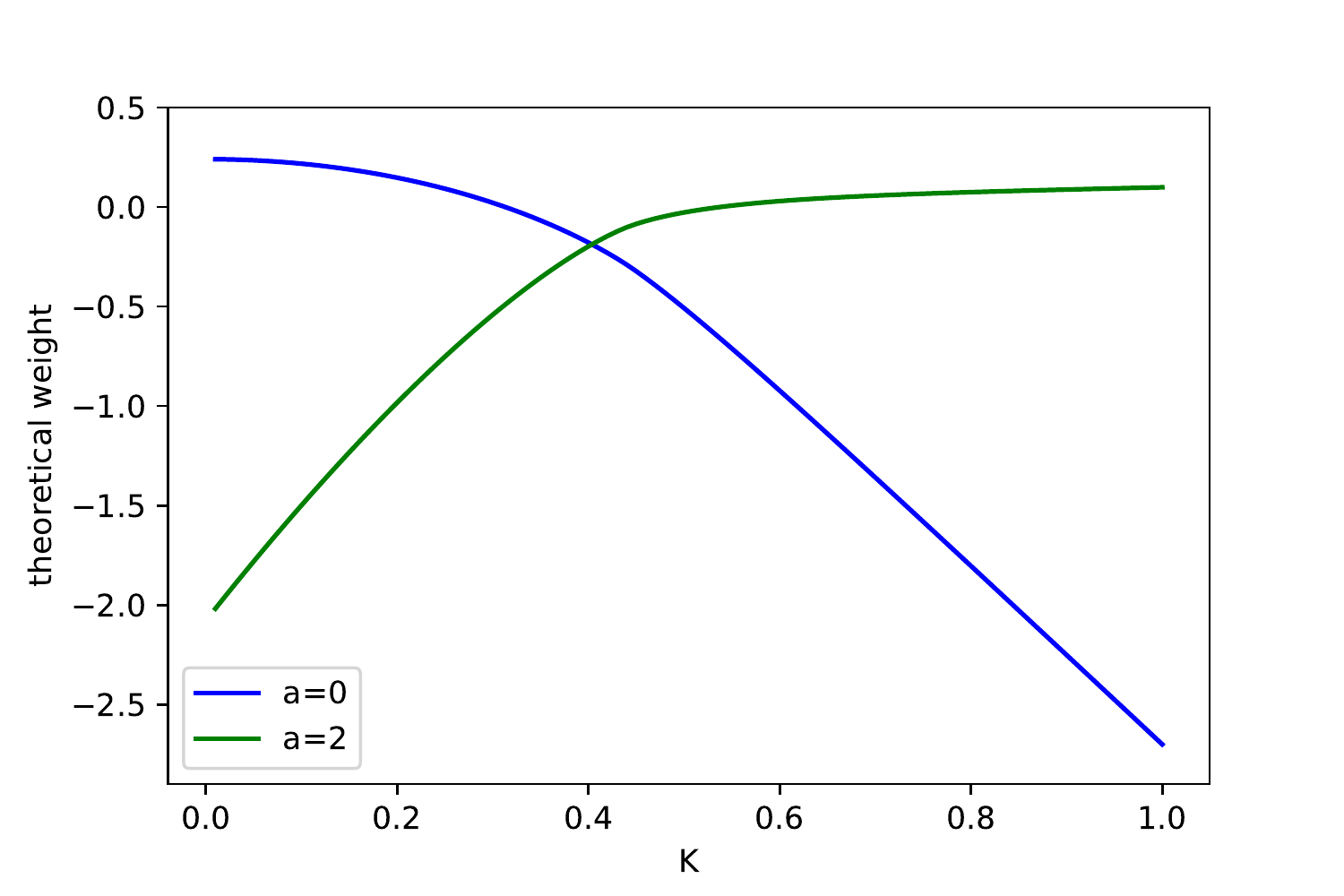}
\caption{Blue and green curves correspond to $w^{(2)}_{K0}$ and $w^{(2)}_{K2}$ in \eqref{eq:theory_weight}, respectively. The horizontal axes represent temperatures $K$ of the input 2D Ising configurations. See the main text for detail.}
\label{fig:theory_weight}
\end{figure} 

Then, $y_K= \text{softmax}(w^{(2)}_{Ka}x_a+b^{(2)}_K)$ with
Eqs.~(\ref{Sec:xa}) and (\ref{eq:theory_weight}) yields the following output;
\begin{align}
 \label{eq:output_CNN}
  y_K &= \text{softmax}(-(p+r)KE-2\epsilon(\phi G(K)+pK)),
\end{align}
where we dropped $K$-independent terms because they do not affect the outcome of the max function.
The expression indeed takes the form of $y_K^\text{theory}$ \eqref{eq:theory_output} if $-2\epsilon(\phi G(K)+pK) = (p+r)F$ is satisfied.
We plot the theoretically predicted weights in Fig.~\ref{fig:theory_weight}:
\begin{align}
w^{(2)}_{K0}= \frac{p+r}{2\epsilon} F(K) + q, 
\quad
w^{(2)}_{K2}= \frac{p+r}{2\epsilon\phi} F(K) + \big(\frac{p}{\phi}+r\big) K + s.
\end{align}
$F(K)$ is the free energy analytically calculated in 2D Ising model.
It agrees very well with the actual weights obtained from the trained neural network shown as blue and green curves in Fig.~\ref{fig:weight_CNN}.
Based on these considerations, we conclude that the information of the
phase transition is again encoded in the fully-connected weight because $G(K)$ in
Eq.~(\ref{eq:theory_weight}) is directly related to the free energy. 
Particularly, the critical temperature is obtained by detecting the enhancement of the second derivative of weights with respect to temperature.

It is noted that the model or the weight parametrization demonstrated above is one of many possibilities yielding the same temperature prediction. For example, the quadratic $K$-dependence could arise from the biases on the last layer instead of the weights~\cite{Aoki2018}. In that case, the information of phase transition or free energy should be encoded in the bias. How the physical information is stored in the machine parameters depends on the architecture of neural network including the number of hidden units or activation functions.

\vspace{1em}

Having gained the insight on the internal structure of the neural networks, we argue on why the convolutional neural network works better than the one composed only of fully-connected layers in predicting temperatures from physical viewpoint.
As we have observed in this section, the latter captures the magnetization in the intermediate layer, while the former extracts the internal energy, based on which it tries to infer the temperature. However, the temperature-dependence of magnetization in the Ising model is small except around the critical temperature, compared with that of internal energy.  Thus, it is plausible that the neural network extracts information of the internal energy for more accurate temperature prediction. Indeed, our convolutional neural network successfully learns the internal energy to achieve higher accuracy. The better performance by the convolutional neural network should be readily understood. It is due to the fact that the convolutional layers identify the feature of input spin configurations by exploiting their spatial structure such as spatial correlation of Ising spins. The fully-connected layers, however, are blind to the structure.

Finally, we mention that the discussion given for 2D Ising model also holds for the 3-state Potts model.

\section{Conclusion}\label{sec:conclusion}

We revisited the phase transition detection of the 2D Ising model based on temperature-supervised machine learning to clarify the underlying mechanism.
\blue{...}

We first demonstrated that the fully-connected neural network shows the drastic change in its weight structure of the second layer as a result of training on 2D Ising model as well as 3-state Potts model.
Closer look at the neural network with 3 hidden units revealed that it actually captures the magnetization in the first layer. The phase transition detection is, however, a consequence of low prediction accuracy of temperature on input configurations except around the critical temperature.

On the other hand, employing the convolutional neural network, we succeeded in improving the temperature-prediction accuracy.
It turned out that the trained convolutional network captures the internal energy of input configurations instead of magnetization.
Also, the weights in the last fully-connected layer do not show any drastic change in their value in contrast to the former case and this aspect is understood from the viewpoint of optimal temperature prediction. 
In this case, the weights are proportional to free energy, and hence, the physical information including the critical temperature is again encoded in them.

Interestingly, the trained neural network extract different physical information depending on how they are trained. A ``bad'' neural network in terms of temperature prediction tries to capture the magnetization of input spin configurations, which happens to be convenient for detecting critical temperature as it is the order parameter. Improvement of the network architecture allow us to construct a ``good (better)'' temperature predictor. But the information of phase transition is encoded in the network more implicitly.


\section*{Acknowledgment}

We thank Akinori Tanaka for providing us with the code used
in~\cite{Tanaka:2016rtu}. We are also grateful to Yutaka Yamaguti for his useful comments.
K.K. is supported by the Grants-in-Aid for Scientific Research
 from JSPS (No. 18K03618).
The work of A.T. was supported by the RIKEN Special Postdoctoral Researcher program.
\appendix

\setcounter{equation}{0}
\section{Temperature prediction and its accuracy}\label{sec:temperature_prediction}

We think about how temperature prediction of Ising spin configuration works~\cite{Aoki2018}.
We start with preparing Ising spin configurations generated by Markov-chain Monte-Carlo algorithm at target temperatures. The configurations at a fixed temperature $K$ are distributed over energy with mean $\langle E\rangle$\footnote{$\langle E\rangle$ stands for a thermal expectation value of the Hamiltonian's eigenvalues.} and variance $\langle E^2-\langle E\rangle^2\rangle$, each of which is given by
\begin{align}
 \label{eq:mean}
 &\langle E\rangle = \frac{\sum_{\{\sigma_i\}}H(\{\sigma_i\})\e^{-K H(\{\sigma_i\})}}{\sum_{\{\sigma_i\}}\e^{-K H(\{\sigma_i\})}}
 =-\frac{\p}{\p K}\ln\calZ,
 \\
 \label{eq:variance}
 &\langle E^2-\langle E\rangle^2\rangle = -\frac{\p}{\p K}\langle E\rangle
 = \frac{\p^2}{\p K^2}\ln\calZ.
\end{align}

\begin{figure}[t]
\centering
\includegraphics[width=0.5\textwidth]{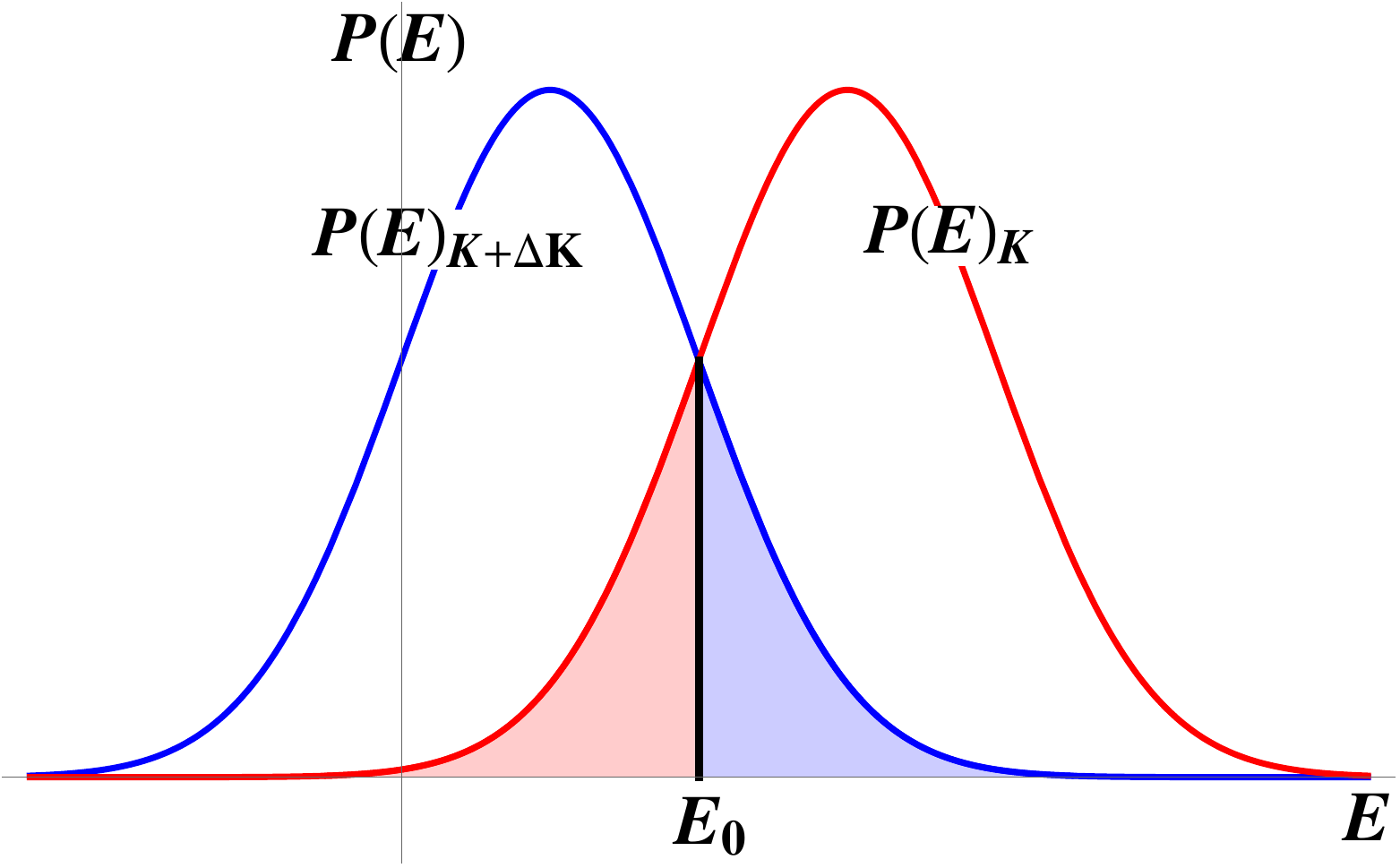}
\caption{Energy probability distributions at two nearby temperatures
 $K$ and $K+\Delta K$, which have overlap shown by red and blue shaded areas. The red distribution is generated at a temperature $K$. The configurations are, however, misclassified as those at temperature $K+\Delta K$ by maximum-like method if they are in the red shaded area because $P(E)_{K+\Delta K}>P(E)_{K}$ below $E_0$. The same argument holds for configurations generated in the blue shaded area.}
\label{fig:overlap}
\end{figure}

Given a spin configuration, we consider the optimal prediction of temperature.
Energy can be calculated for each configuration. 
It is noted that, since the standard deviation of energy density $E/L^2$ is proportional to $L^{-1}$, the energy probability distribution does not admit width in infinite system. In such case, provided a certain configuration, we should be able to correctly predict the temperature at which it is generated by calculating its energy density. 
However, the distribution at each temperature has finite width if we consider finite systems. In that case, temperature prediction does not necessarily give the correct answer because there are overlaps between energy distributions (Fig.~\ref{fig:overlap}).
The best we can do is to guess the temperature of the configuration by maximally likelihood estimate, namely, the temperature that is most likely to yield the configuration's energy is the optimally predicted temperature.
Consequently, there is an upper bound in the accuracy of prediction
from the overlap of the energy probability distributions in finite system. 

\begin{figure}[t]
\centering
\includegraphics[width=0.6\textwidth]{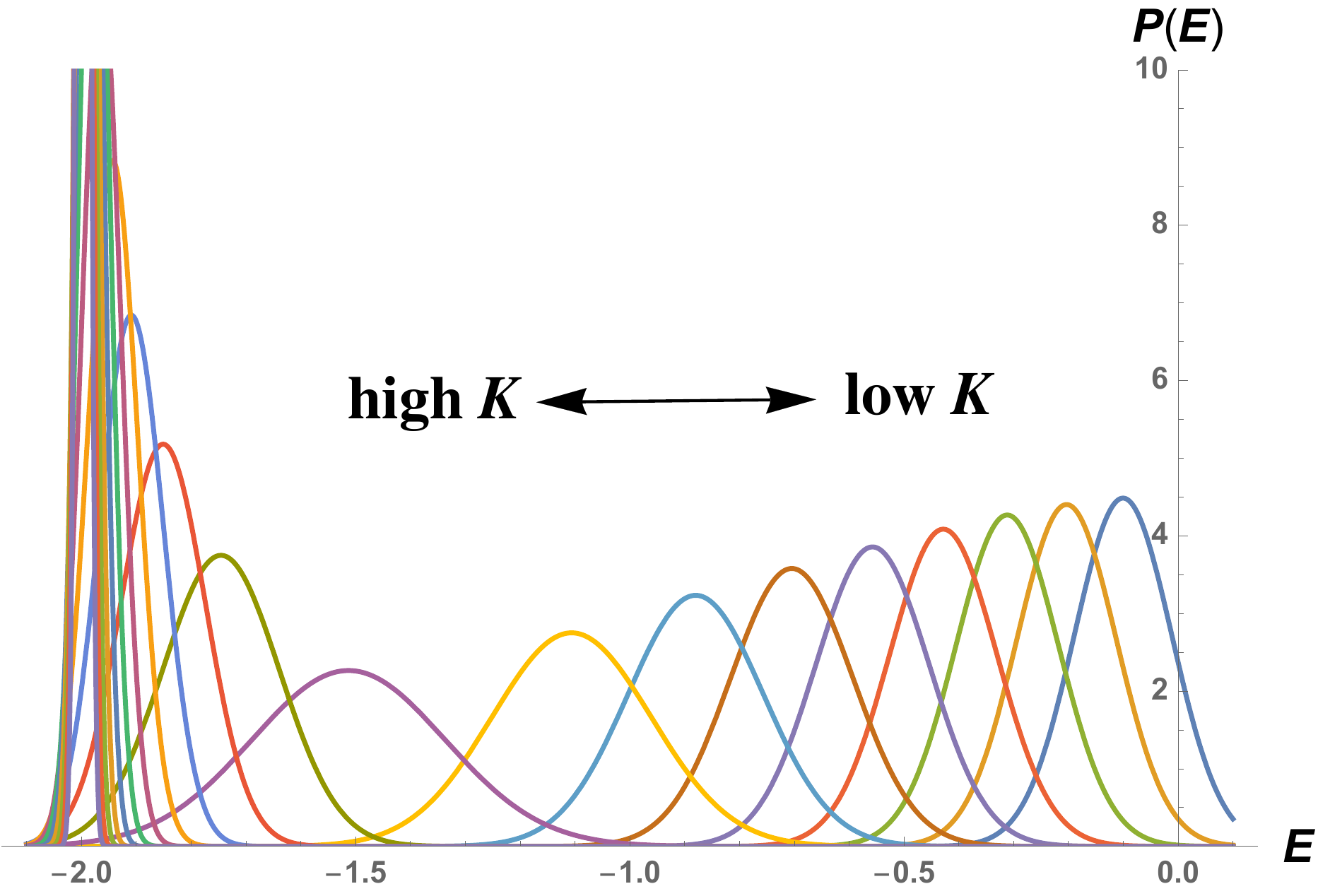}
\caption{Energy probability distributions in two dimensional Ising model at temperatures $K=0.05, 0.1, 0.15,\dots,1.0$. The vertical axis is energy per site and number of sites is taken to be $16\times 16$.}
\label{fig:2DIsingDist}
\end{figure}

Let us take a look at two dimensional Ising model in more detail. Since it is exactly solved~\cite{onsager1944crystal}, we have an explicit expression of the free energy density,
\begin{align}
 f = -\frac{\ln\calZ}{N} 
 = -\frac{\ln 2}{2} -\ln[\cosh(2K)] 
 -\frac{1}{2\pi}\int_0^\pi\diff\theta \ln\left[1+\sqrt{1-\left(\frac{2\sinh(2K)}{\cosh^2(2K)}\cos\theta\right)^2}\right],
\end{align}
with the number of sites $N$. We do not incorporating the finite size effect.
Then, we obtain the energy probability distributions approximated by the Gaussian distributions with mean $\langle E\rangle$ and variance $\langle(E-\langle E\rangle)^2\rangle$ at the temperatures of our interest.
We show energy distributions at $20$ different temperatures $K=0.05, 0.1, 0.15,\dots,1.0$ in Fig.~\ref{fig:2DIsingDist}.
We see that the overlaps between distributions are significant while they are reduced around the critical temperature $K_\text{c}^\text{exact}\sim 0.4407$. The accuracy obtained by maximum-likelihood estimate is as low as 40.1\%, implying that a thermometer constructed by machine learning can achieve an accuracy 40.1\% at most.
Nevertheless, it turns out that the trained neural network can work as a phase transition detector.

\bibliographystyle{JHEP}
\bibliography{ML}

\providecommand{\href}[2]{#2}\begingroup\raggedright\begin{thebibliography}{10}

\bibitem{landau1937theory}
L.~D. Landau, \emph{{On the theory of phase transitions. I.}}, {\emph{Zh. Eksp.
  Teor. Fiz.} {\bf 11} (1937) 19}.

\bibitem{ginzburg1950theory}
V.~L. Ginzburg and L.~D. Landau, \emph{On the theory of superconductivity},
  {\emph{Zh. eksp. teor. Fiz} {\bf 20} (1950) 35}.

\bibitem{carrasquilla2017machine}
J.~Carrasquilla and R.~G. Melko, \emph{{Machine learning phases of matter}},
  \href{http://dx.doi.org/10.1038/nphys4035}{\emph{Nature Physics} {\bf 13}
  (2017) 431}.

\bibitem{Tanaka:2016rtu}
A.~Tanaka and A.~Tomiya, \emph{{Detection of phase transition via convolutional
  neural network}}, \href{http://dx.doi.org/10.7566/JPSJ.86.063001}{\emph{J.
  Phys. Soc. Jap.} {\bf 86} (2017) 063001}.

\bibitem{ohtsuki2016deep}
T.~Ohtsuki and T.~Ohtsuki, \emph{{Deep learning the quantum phase transitions
  in random two-dimensional electron systems}},
  \href{http://dx.doi.org/10.7566/JPSJ.85.123706}{\emph{Journal of the Physical
  Society of Japan} {\bf 85} (2016) 123706}.

\bibitem{mano2017phase}
T.~Mano and T.~Ohtsuki, \emph{{Phase Diagrams of Three-Dimensional Anderson and
  Quantum Percolation Models Using Deep Three-Dimensional Convolutional Neural
  Network}}, \href{http://dx.doi.org/10.7566/JPSJ.86.113704}{\emph{Journal of
  the Physical Society of Japan} {\bf 86} (2017) 113704}.

\bibitem{schindler2017probing}
F.~Schindler, N.~Regnault and T.~Neupert, \emph{{Probing many-body localization
  with neural networks}},
  \href{http://dx.doi.org/10.1103/PhysRevB.95.245134}{\emph{Physical Review B}
  {\bf 95} (2017) 245134}.

\bibitem{broecker2017machine}
P.~Broecker, J.~Carrasquilla, R.~G. Melko and S.~Trebst, \emph{{Machine
  learning quantum phases of matter beyond the fermion sign problem}},
  \href{http://dx.doi.org/10.1038/s41598-017-09098-0}{\emph{Scientific reports}
  {\bf 7} (2017) 8823}.

\bibitem{ch2017machine}
K.~Ch'ng, J.~Carrasquilla, R.~G. Melko and E.~Khatami, \emph{{Machine learning
  phases of strongly correlated fermions}},
  \href{http://dx.doi.org/10.1103/PhysRevX.7.031038}{\emph{Physical Review X}
  {\bf 7} (2017) 031038}.

\bibitem{zhang2017quantum}
Y.~Zhang and E.-A. Kim, \emph{{Quantum loop topography for machine learning}},
  \href{http://dx.doi.org/10.1103/PhysRevLett.118.216401}{\emph{Physical review
  letters} {\bf 118} (2017) 216401}.

\bibitem{ponte2017kernel}
P.~Ponte and R.~G. Melko, \emph{{Kernel methods for interpretable machine
  learning of order parameters}},
  \href{http://dx.doi.org/10.1103/PhysRevB.96.205146}{\emph{Physical Review B}
  {\bf 96} (2017) 205146}.

\bibitem{zhang2017machine}
Y.~Zhang, R.~G. Melko and E.-A. Kim, \emph{{Machine learning $\mathbb{Z}_2$
  quantum spin liquids with quasiparticle statistics}},
  \href{http://dx.doi.org/10.1103/PhysRevB.97.045207}{\emph{Physical Review B}
  {\bf 96} (2017) 245119}.

\bibitem{arai2018deep}
S.~Arai, M.~Ohzeki and K.~Tanaka, \emph{{Deep Neural Network Detects Quantum
  Phase Transition}},
  \href{http://dx.doi.org/10.7566/JPSJ.87.033001}{\emph{Journal of the Physical
  Society of Japan} {\bf 87} (2018) 033001}.

\bibitem{beach2018machine}
M.~J.~S. Beach, A.~Golubeva and R.~G. Melko, \emph{{Machine learning vortices
  at the Kosterlitz-Thouless transition}},
  \href{http://dx.doi.org/10.1103/PhysRevB.97.045207}{\emph{Phys. Rev. B} {\bf
  97} (Jan, 2018) 045207}.

\bibitem{Wetzel:2017ooo}
S.~J. Wetzel and M.~Scherzer, \emph{{Machine Learning of Explicit Order
  Parameters: From the Ising Model to SU(2) Lattice Gauge Theory}},
  \href{http://dx.doi.org/10.1103/PhysRevB.96.184410}{\emph{Phys. Rev.} {\bf
  B96} (2017) 184410}.

\bibitem{richter2018machine}
M.~Richter-Laskowska, H.~Khan, N.~Trivedi and M.~Ma{\'s}ka, \emph{{A machine
  learning approach to the Berezinskii-Kosterlitz-Thouless transition in
  classical and quantum models}},
  \href{http://arxiv.org/abs/arXiv:1809.09927}{{\tt arXiv:1809.09927}}.

\bibitem{Suchsland2018}
P.~Suchsland and S.~Wessel, \emph{Parameter diagnostics of phases and phase
  transition learning by neural networks},
  \href{http://dx.doi.org/10.1103/PhysRevB.97.174435}{\emph{Phys. Rev. B} {\bf
  97} (2018) 174435}.

\bibitem{Kim2018}
D.~Kim and D.-H. Kim, \emph{Smallest neural network to learn the ising
  criticality}, \href{http://dx.doi.org/10.1103/PhysRevE.98.022138}{\emph{Phys.
  Rev. E} {\bf 98} (Aug, 2018) 022138}.

\bibitem{wang2016discovering}
L.~Wang, \emph{{Discovering phase transitions with unsupervised learning}},
  \href{http://dx.doi.org/10.1103/PhysRevB.94.195105}{\emph{Physical Review B}
  {\bf 94} (2016) 195105}.

\bibitem{morningstar2017deep}
A.~Morningstar and R.~G. Melko, \emph{{Deep learning the Ising model near
  criticality}},  \href{http://arxiv.org/abs/arXiv:1708.04622}{{\tt
  arXiv:1708.04622}}.

\bibitem{van2017learning}
E.~P. Van~Nieuwenburg, Y.-H. Liu and S.~D. Huber, \emph{{Learning phase
  transitions by confusion}},
  \href{http://dx.doi.org/10.1038/nphys4037}{\emph{Nature Physics} {\bf 13}
  (2017) 435}.

\bibitem{hu2017discovering}
W.~Hu, R.~R. Singh and R.~T. Scalettar, \emph{{Discovering phases, phase
  transitions, and crossovers through unsupervised machine learning: A critical
  examination}},
  \href{http://dx.doi.org/10.1103/PhysRevE.95.062122}{\emph{Physical Review E}
  {\bf 95} (2017) 062122}.

\bibitem{wetzel2017unsupervised}
S.~J. Wetzel, \emph{{Unsupervised learning of phase transitions: From principal
  component analysis to variational autoencoders}},
  \href{http://dx.doi.org/10.1103/PhysRevE.96.022140}{\emph{Physical Review E}
  {\bf 96} (2017) 022140}.

\bibitem{Iso:2018yqu}
S.~Iso, S.~Shiba and S.~Yokoo, \emph{{Scale-invariant Feature Extraction of
  Neural Network and Renormalization Group Flow}},
  \href{http://dx.doi.org/10.1103/PhysRevE.97.053304}{\emph{Phys. Rev.} {\bf
  E97} (2018) 053304}.

\bibitem{Zhang2018}
P.~Zhang, H.~Shen and H.~Zhai, \emph{Machine learning topological invariants
  with neural networks},
  \href{http://dx.doi.org/10.1103/PhysRevLett.120.066401}{\emph{Phys. Rev.
  Lett.} {\bf 120} (2018) 066401}.

\bibitem{Sun2018}
N.~Sun, J.~Yi, P.~Zhang, H.~Shen and H.~Zhai, \emph{Deep learning topological
  invariants of band insulators},
  \href{http://dx.doi.org/10.1103/PhysRevB.98.085402}{\emph{Phys. Rev. B} {\bf
  98} (2018) 085402}.

\bibitem{kingma2014adam}
D.~P. Kingma and J.~Ba, \emph{Adam: A method for stochastic optimization},
  \href{http://arxiv.org/abs/arXiv:1412.6980}{{\tt arXiv:1412.6980}}.

\bibitem{onsager1944crystal}
L.~Onsager, \emph{{Crystal Statistics. I. A Two-Dimensional Model with an
  Order-Disorder Transition}},
  \href{http://dx.doi.org/10.1103/PhysRev.65.117}{\emph{Phys. Rev.} {\bf 65}
  (1944) 117--149}.

\bibitem{Wu1982}
F.~Y. Wu, \emph{The potts model},
  \href{http://dx.doi.org/10.1103/RevModPhys.54.235}{\emph{Rev. Mod. Phys.}
  {\bf 54} (1982) 235--268}.

\bibitem{baxter1973potts}
R.~J. Baxter, \emph{{Potts model at the critical temperature}},
  \href{http://dx.doi.org/10.1088/0022-3719/6/23/005}{\emph{Journal of Physics
  C: Solid State Physics} {\bf 6} (1973) L445}.

\bibitem{alexander1975s}
S.~Alexander and D.~J. Amit, \emph{{When the Landau criterion fails
  qualitatively}},
  \href{http://dx.doi.org/10.1088/0305-4470/8/12/015}{\emph{Journal of Physics
  A: Mathematical and General} {\bf 8} (1975) 1988}.

\bibitem{Aoki2018}
K.~Aoki, T.~Fujita and T.~Kobayashi, \emph{What does deep learning of
  statistical system learn?}, {\emph{Journal of the Japanese Society for
  Artificial Intelligence} {\bf 33} (2018) }.

\end{thebibliography}\endgroup

\end{document}